\documentclass[aps,preprint,amssymb,amsmath]{revtex4-1}
\usepackage{graphicx}
\usepackage{epstopdf}

\begin{document}

\title
{Seismic quiescence and b-value decrease before large events in forest-fire model}

\author{Tetsuya Mitsudo}
\email{mitsudo@scphys.kyoto-u.ac.jp}
\affiliation{Department of Physics, Kyoto University, 
Kitashirakawa-Oiwakecho, Sakyo-ku, Kyoto 606-8502, Japan}
\author{Takahiro Hatano}
\author{Naoyuki Kato}
\affiliation{Earthquake Research Institute, The University of Tokyo,
1-1-1 Yayoi, Bunkyo, Tokyo 113-0032, Japan}

\date{\today}

\begin{abstract}
Forest fire models may be interpreted as a simple model for earthquake occurrence
by translating trees and fire into stressed segments of a fault and their rupture, respectively.
Here we adopt a two-dimensional forest-fire model in continuous time, 
and focus on the temporal changes of seismicity and the $b$-value.
We find the $b$-value change and seismic quiescence prior to large earthquakes 
by stacking many sequences towards large earthquakes.
As the magnitude-frequency relation in this model is directly related to
the cluster-size distribution, decrease of the $b$-value can be explained 
in terms of the change in the cluster-size distribution.
Decrease of the $b$-value means that small clusters of stressed sites
aggregate into a larger cluster.
Seismic quiescence may be attributed to the decrease of stressed sites
 that do not belong to percolated clusters.
\end{abstract}


\maketitle

\section{Introduction}
Although seismic quiescence prior to large earthquakes has been reported in some studies \cite{Mog,Kat},
it is important to note that seismic quiescence is not always observed.
On the contrary, sometimes seismicity appears to be active prior to large earthquakes \cite{Run}.
To utilize seismic quiescence as a precursory index of a large earthquake,
the physical mechanism of such anomalies in seismicity must be clarified
in the context of the preparation process of large earthquakes \cite{Wys,Sch88,Mai91,Kato}.

Another potential index is the $b$-value decrease, which is often reported for major earthquakes 
\cite{Imo,Nanjo}.
Again, however, the $b$-value decrease does not necessarily mean a large earthquake in the near future.
As a result, the $b$-value decrease can be regarded as a precursor of a large earthquake only after the occurrence of a large earthquake.
At the same time, however, one should pay due attention to the work of \cite{Smi}, who reported the opposite tendency.
Therefore the situation about real earthquakes is rather puzzling.

On the other hand, decrease of the $b$-value prior to a major rupture has been ubiquitously observed in rock fracture experiments \cite{Sch68}.
This may be interpreted as negative dependence of the $b$-value on differential stress, and has been indirectly supported by some observational studies on the $b$-value with respect to the faulting type \cite{Schr}, to the depth \cite{Mori1997,Spada}, and to the plate age \cite{Nishikawa}.
These observational and experimental data are reinterpreted from the viewpoint of stress dependence by \cite{Sch15}.
At the same time, however, one must pay serious attention to an experiment of \cite{Mogi},
in which the $b$-value decreases as a function of time at a {\it constant stress level}.
In this paper, Mogi claimed that complexity in the fabric of microcracks, or the heterogeneity of the stress field, may be more important than the differential stress itself.

Because earthquakes involve numerous physical processes that are spanned in a wide range of
 spatiotemporal scales, identification of dominant physical processes behind the $b$-value decrease
and seismic quiescence are not straightforward.
In addition, the verification of any hypothesis on large earthquakes may be problematic due to the lack of sufficient number of samples that ensures statistical significance.
Inaccessibility to physical quantities in earthquake faults makes the problem even more difficult.

Along the line of thought, to complement observational and experimental studies,
simple physical models that reproduce seismicity may be helpful
in guessing potential mechanisms for the changes in seismicity prior to large events
\cite{Kato,Hai00,Kaw}.
In addition, accessibility to all the physical quantities and statistical significance are ensured.

Among such models, we adopt a probabilistic cellular automaton of forest fire \cite{Bak90,Dro92},
which simulates planting and burning of trees in a lattice system.
This model may be also viewed as a simple model for earthquakes \cite{Tur99-1,Tur99-2,Tej,Jag}.
Planting of a tree corresponds to loading a fault segment with stress.
The ignition and the spread of fire correspond to the triggering of an earthquake and the rupture propagation, respectively.
Using this simple model, we show that quiescence and change in the $b$-value occur before large earthquakes.

\section{Model and Method}
The present model is defined on a square grid with $L\times L$ sites.
The internal states of each site are binary: the stressed state and the unstressed state.
The stressed state is ready for an earthquake to be triggered.
The neighboring stressed sites are expressed as ``connected'', and an assembly of connected sites is expressed as a ``cluster''.
Initially, all the sites are unstressed.
(The choice of the initial condition does not affect the long-time behavior of the model.)

Here we adopt an algorithm for continuous time \cite{He93,Sch}.
The dynamics is described as follows.
A site is chosen randomly.
If the site is unstressed, the site is stressed with the probability of $p\Delta t$.
Here $p$ is the stressing rate and $\Delta t$ is a small time step.
If the site is stressed, an earthquake is triggered with the probability of $f\Delta t$.
Here $f$ is the triggering rate.
Importantly, once an earthquake is triggered on the initial site, the all sites 
that belong to the same cluster rupture immediately.
Then the all ruptured sites are unstressed.
This process represents an earthquake in the present model, and the number of ruptured sites defines the size of earthquake.
Throughout this paper, the size-frequency relation is denoted by $N(s)$.

After the occurrence of an earthquake, another site is chosen randomly, and the same procedure applies.
This process is repeated $L^2$ times in one time step, $\Delta t$, namely the unit time contains $L^2/\Delta t$ loops of random choice of a site and stressing/triggering.
We assume that the time scale of loading is much longer than that of the rupture propagation.
Stressing and unstressing are regarded as an instantaneous process.
A schematic of time evolution of the present model is shown in FIG.~1.

Note that we have only two control parameters: $f/p$ and $L$.
We adopt $L=256$, $p=1$, and $\Delta t=0.01$ in the present simulations, and therefore the only relevant parameter here is $f$.
We confirm that the system behavior is roughly determined by the value of $f/L^2$, rather than $f$ itself.
To ensure statistical significance, any statistical quantity is calculated in the duration of $2^{20}$ (in the simulation time unit) for each given value of the parameter $f$.

A closed boundary condition is adopted, where the rupture does not propagate to outside of the system.
The Hoshen-Kopelman algorithm is used for the identification of the clusters \cite{Hsk,Aha}.
Note that there are several other algorithms for simulating forest-fire models
\cite{Bak90,Dro92,He93,Tur99-1,Tej,Jag}.
Although subtle statistical properties of the model may depend on numerical algorithm,
we do not consider such algorithm-dependence in this paper.

Throughout this study, a large earthquake is defined as the unstressing process of a percolated cluster \cite{New}.
A percolated cluster spans from one edge to the other edge, either vertically or horizontally.
Therefore, the size of any percolated earthquake must be larger than $L$.

\section{Main Results}
Throughout this paper, we investigate three representative values of $f$: 
$f =100/L^2$, $10/L^2$, $1/L^2$.
Triggering occurs most frequently at $f=100/L^2$, and therefore 
stressed sites cannot significantly increase in comparison with the other two cases.
We define the stressed fraction, $\rho$, as the ratio of the number of 
stressed sites to the total site number ($L^2$).
The stressed fraction is smaller at higher triggering rates, and the average value just before a large event is approximately $0.4$ ($f=100/L^2$), $0.6$ ($f=10/L^2$), and $0.8$ ($f =1/L^2$), respectively.
As we explain later, the stressed fraction may be interpreted as the average stress in the system and therefore an important quantity throughout this study.

It is well known that the size-frequency relations in forest-fire models
 are well approximated by power laws \cite{Tur99-1,Tur99-2,Tej},
which resembles the Gutenberg-Richter (GR) law for real earthquakes.
We show that the present model also reproduces the GR law.
In FIG.~2, the size-frequency relations $N(s)$ at the three values of $f/L^2$ are shown.
In the small earthquake range (approximately $s<5000$), the GR law 
well approximates the observed size-frequency relations irrespective of the values of $f$.
The exponent is approximately $-1.2$ for each value of $f$.
This value is also observed in other variations of forest-fire models \cite{Gra,Tur99-1}.
Due to the definition of event size, this exponent cannot be directly compared to 
the $b$ value for the GR law.
In the large earthquake range (about $s>5000$), in which earthquakes 
are mostly percolated earthquakes, the size-frequency relations exhibit
upward deviation from the GR law.
In particular, a clear peak is observed for $f=10/L^2$ and $f=1/L^2$ cases.
This peak may be regarded as characteristic earthquakes.
The size of characteristic earthquakes are roughly $s\sim40000$ for $f=10/L^2$
 and $s\sim L^2$ for $f=1/L^2$.
On the other hand, the size-frequency relation decays at much smaller size
($s\sim 10000$) for $f=100/L^2$.
Similar deviations from the GR law are widely observed in the studies of earthquakes \cite{Mai},
a forest fire model \cite{Tej}, and some related discrete models \cite{Kaw,Mor}.
If we were to adopt their classification, the case of $f=1/L^2$ and $10/L^2$ may be 
referred to as the supercritical regime, whereas $f=100/L^2$ the subcritical regime.

Importantly, the size-frequency relation in the present model is directly related to
the cluster size distribution.
For an earthquake of size $s$ to occur, a stressed site that belongs to a cluster
 of size $s$ must be chosen.
This probability is proportional to $s n_c(s)$, where $n_c(s)$ is the number of
clusters of size $s$.
Therefore, the probability of an earthquake of size $s$ is proportional to $fsn_c(s)$;
i.e., $N(s)\propto sn_c(s)$.
We confirm this relation numerically.

To observe quiescence prior to percolated earthquakes, 
the number of earthquakes per unit time (i.e., the event rate) is monitored.
To do this, the number of earthquakes in each $\Delta t$ bin is recorded
for certain duration prior to a percolated earthquake.
The event rate is regarded as a function of time before a percolated earthquake, $\tau_p$.
A percolated earthquake occurs at $\tau_p=0$, and the event rate is monitored for $\tau_p\le 0$.
One sequence of events contains one percolated earthquake.
To ensure significant statistics, sequences of earthquakes are
 averaged over many percolated earthquakes: $2.33\times 10^5$ for $f=100/L^2$,
$1.04\times 10^6$ for $f=10/L^2$ and $4.88\times 10^5$ for $f=1/L^2$.
We need $2^{24}$ time steps for such large numbers of percolated earthquakes.
The time before a percolated earthquake $\tau_p$ is normalized by
 the mean recurrence time of percolated earthquakes, ${\bar T}$:
 $\bar{T}=4.51$ for $f=100/L^2$, $\bar{T}=1.00$ for $f=10/L^2$, and $\bar{T}=2.14$ for $f=1/L^2$.
Interestingly, the mean recurrence time ${\bar T}$ is not monotonic with respect to $f$.
If $f$ is too large, unstressing overwhelms stressing and therefore
 the growth of large clusters is suppressed.
If $f$ is too small, stressing overwhelms unstressing and therefore
 the event rate itself is suppressed, and so is the rate of large events.
The event rate defined as above is shown in FIG.~3.
Quiescence before percolated earthquakes is apparent in all cases,
but the duration depends on the parameter, $f$.
The average duration of quiescence is $0.3$ to $0.4$ for $f=10/L^2$ and $1/L^2$, and $0.08$ for $f=100/L^2$.

To see the mechanism of quiescence, it is useful to introduce another stressed fraction,
$\rho_{\rm wo}$, in addition to the stressed fraction $\rho$.
It is defined by disregarding the stressed sites that constitute percolated clusters.
$\rho_{\rm wo}$ is the ratio of the number of stressed sites
that do not constitute a percolated cluster to the total site number.
Obviously, the deviation between the two quantities indicates the emergence of percolated clusters.
Recalling that the event rate is proportional to $f\rho$,
the rate of earthquakes other than percolated one is proportional to $f\rho_{\rm wo}$.
Therefore, the decrease of $\rho_{\rm wo}$ is equivalent to the decrease of smaller earthquakes;
 namely, quiescence.
In FIG.~4, the temporal behaviors of $\rho$ and $\rho_{\rm wo}$ are shown.
As in FIG.~3, the both quantities are averaged for many percolated earthquakes.
For the three values of $f$ shown here, $\rho_{\rm wo}$ decreases before a percolated earthquake.
The beginning of the decrease of $\rho_{\rm wo}$ is $\tau_p/\bar{T}\sim -0.4$ to $-0.3$ for $f=10/L^2$ and $1/L^2$,
 and $\tau_p/\bar{T}\sim -0.1$ for $f=100/L^2$, respectively.
This approximately corresponds to the beginning of quiescence shown in FIG.~3.
Therefore, we can conclude that quiescence before percolated earthquakes 
is due to the decrease of stressed sites that do not constitute percolated clusters.

On the other hand, as $\rho$ increases during the entire period,
the total number of stressed sites also keeps increasing.
However, triggering of stressed sites that constitute percolated clusters must be
 avoided before a percolated earthquake.
If a stressed site that belongs to a percolated cluster were triggered, 
a percolated earthquake would occur at $\tau_p\le0$.

One can also aware the oscillation of stressed fraction for $f=100/L^2$.
This implies that relatively large events tend to occur periodically before percolated earthquake.
This may be an interesting precursor of percolated earthquake, but we do not describe it in detail here.

Another precursory index for large earthquakes is change of the $b$-value.
To observe the temporal change of the $b$-value, the time window of $-1\leq\tau_p/\bar{T}\leq 0$ is divided into $16$ bins.
The size-frequency relation is calculated in each bin.
The $b$-value is determined for each size-frequency relation by applying
 the least-square method in log-log scale in the range $10 \leq s\leq 1000$,
 in which the GR law well approximates the size-frequency relation.
Temporal changes of the $b$-value defined in this manner are presented in FIG.~5(A).
In the case of $f=1/L^2$, the $b$-value decreases drastically from $\tau_p/{\bar T}\simeq -0.2$ toward percolated earthquake.
For $f=10/L^2$, the $b$-value decreases in the latter period, $-0.5\le\tau_p/\bar{T}\le-0.1$.
Interestingly, however, the $b$-value then turns to increase just prior to percolated earthquake,
$-0.1\le\tau_p/{\bar T}\le 0$.
The change of the $b$-value is not visible for $f=100/L^2$.

These temporal changes of the $b$-value are partially explained if one recalls that the size-frequency relation is directly related to the cluster-size distribution: $N(s)\propto sn_c(s)$.
The GR law holds when and only when the cluster-size distribution is a power law,
and the exponent for the cluster-size distribution is $b+1$.
Therefore, decrease of the $b$-value implies relative increase of larger clusters.
Larger clusters are mostly produced by connection of two (or more) preexisting clusters,
which can occur when and only when they are separated by only one-site distance.
Therefore, for this aggregation process to be frequent, namely, for the $b$-value to decrease, 
the stressed fraction $\rho$ must be high enough.

In the case of $f=1/L^2$, the $b$-value decreases for $\tau_p/{\bar T}\ge-0.4$, 
when the stressed fraction $\rho$ exceeds approximately $0.6$ as shown in FIG.~4.
For $\tau_p/{\bar T}\le-0.4$, the $b$-value is virtually constant because the stressed fraction $\rho$ is not sufficiently high for aggregation of large clusters.
One may also notice that the stressed fraction increases gradually in this period and therefore the system is not in a steady state despite the constant $b$-value.
On the other hand, in the early period ($-1\le\tau_p/{\bar T}\le-0.8$), the $b$-value cannot be defined because the GR law does not hold.
This is mainly due to the large variance in the recurrence time of percolated earthquakes.

In the case of $f=10/L^2$, the $b$-value decreases for $\tau_p/{\bar T}\ge-0.5$,
when the stressed fraction $\rho$ exceeds approximately $0.45$.
When the stressed fraction $\rho$ is smaller than $0.45$, the $b$-value increases with time.
This may be attributed to the relative decrease of larger clusters due to the high triggering rate.
However, we do not have clear explanation on the increase of the $b$-value just before percolated earthquake.

In the case of $f=100/L^2$, the triggering is so frequent that the stressed fraction cannot be large enough for further growth of clusters.
In addition, as shown in FIG.~4, the temporal change in $\rho$ is the smallest among the three cases, and therefore the system may be regarded as a steady state to have the time-independent $b$-value.

Importantly, the stressed fraction $\rho$ may approximate the average stress of the entire system.
To see this, let us assume that the shear stress on any stressed site is $\tau_1$ and the residual stress on unstressed sites $\tau_2$.
Then the average stress of the system is $\tau_2+\rho (\tau_1-\tau_2)$.
Therefore, increase of the stressed fraction $\rho$ may be interpreted as increase of the average stress of the system.
Particularly, if a stressed site releases the stress completely upon earthquake, $\tau_2$ is set to be $0$, and the stressed fraction $\rho$ itself is proportional to the average stress.
To show the relation between the stress and the $b$-value explicitly,
the $b$-value in FIG.~5(A) and the stressed fraction $\rho$ in FIG.~4
are replotted in FIG.~5(B).
Although there is no unified curve for the stress dependence of the $b$-value, the apparent trend is negative dependence except for the data at $f=10/L^2$ in $\rho\ge0.6$.
This trend is consistent with the claim that the $b$-value has negative correlation with the differential stress \cite{Sch68,Schr,Sch15} if we admit the correspondence between the stressed fraction and the stress.
In the present model, the trend is opposite for lower stressed fraction, $\rho\le0.4$.
However, we believe that this branch is irrelevant to rock fracture and earthquakes, because there may be no event at very low differential stress.

\section{Discussions and Conclusions}

In this report, quiescence and decrease of the $b$-value prior to
large earthquakes is observed in the forest-fire model.
Quiescence in our model is due to dominance of percolated clusters and relative decrease of small clusters.
This is confirmed by observation of the temporal change of the stressed fraction
and the one without the contribution of percolated clusters.

As the event-size distribution interrelates with the cluster-size distribution in the forest-fire model, decrease of the $b$-value is a consequence of relative decrease of smaller clusters.
Thus, it is not surprising that both the $b$-value decrease and quiescence are observed before large earthquakes.
The essential physical mechanism of quiescence and decrease of the $b$-value 
in the forest-fire model thus involves time evolution of the cluster-size distribution.

Time evolution of the cluster-size distribution consists of the growth/death processes of clusters.
Death of a cluster is an earthquake, and growth of clusters is due to stressing.
Stressing of a site may result in either formation of a single stressed site or union of two preexisting stressed clusters.
The latter is an essential process of cluster growth.
One can model the growth/death processes and write down a time-evolution
law for the cluster-size distribution.
We may analyze the $b$-value change more
quantitatively with the help of such an equation.

We are aware that the forest-fire model may be too simple to be relevant to earthquake faults.
Each cell may represent a segment of a non-planar fault, but a regular square lattice is obviously a gross simplification.
Nevertheless, we believe that the result presented here is still somewhat suggestive.
Particularly, spatial structure of the stressed/unstressed sites may represent the stress heterogeneity on a non-planar fault, which plays an important role in the earthquake complexity.
For instance, the cluster-size distribution in the model should be compared to the stress distribution on a fault: the mean cluster size here should correspond to the correlation length of the heterogeneous stress field.
If the stress distribution is power law, the exponent may have correlation with the $b$-value.
Such a correspondence with the continuum mechanics is an important problem to be addressed as the next step of the present work.

\begin{figure}
\begin{center}
\includegraphics[scale=0.6]{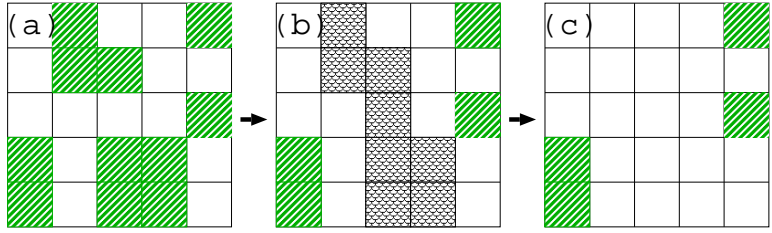}
\caption{An example of time evolution of the present model for $L=5$.
A site filled with green slash represents a stressed site, and the blank sites are unstressed. 
Sites with the black web form a percolated cluster.
There are five clusters panel (a): two clusters of size $1$, 
and clusters of sizes $2,3$ and $4$, respectively.
The site in the middle is stressed in the transition from (a) to (b).
As a result, the cluster of size $3$ and the cluster of size $4$ are connected 
to form a percolated cluster.
The triggering of the percolated cluster occurs in the transition from (b) to (c).
The size of the earthquake is $8$, because the black cluster consists of $8$ loaded sites.
}
\end{center}
\end{figure}
\begin{figure}
 \begin{center}
\includegraphics[scale=0.35]{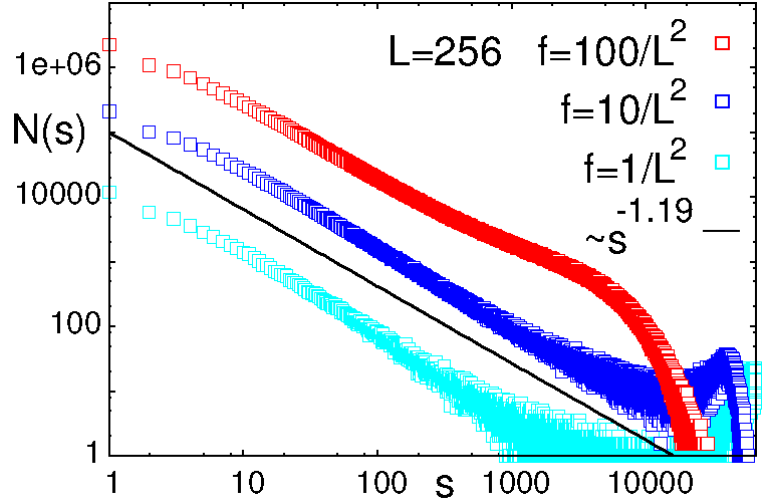}
\caption{The size distribution of number of earthquakes occurred in a
  simulation time $2^{20}$ for $f=100/L^2$(red) $10/L^2$(blue) and
  $1/L^2$(light blue). The solid line is a guideline of the exponent
  $-1.19$.}
 \end{center}
\end{figure}
\begin{figure}
 \begin{center}
\includegraphics[scale=0.5]{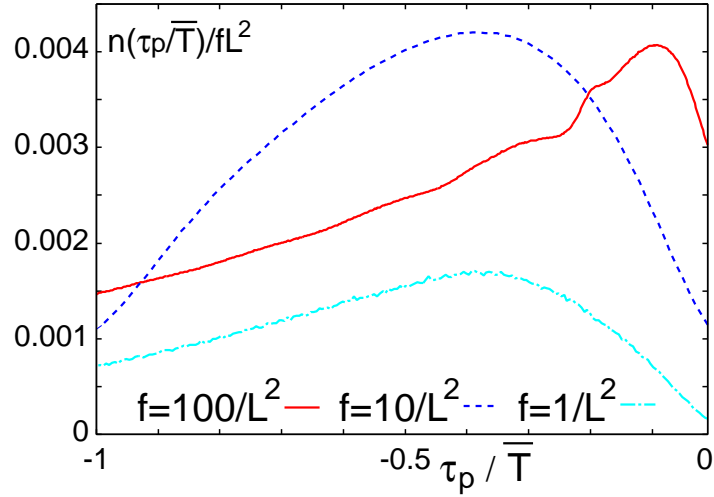}
\caption{Temporal changes in the number of earthquakes $n(\tau_p)L^2/f$
  before the percolated earthquakes for $f=100/L^2$(red),
  $f=10/L^2$(blue) and $f=1/L^2$(light blue). The event rate
  $n(\tau_p/\bar{T})$ is scaled by $f/L^2$ for convenience.}
 \end{center}
\end{figure}
\begin{figure}
\begin{center}
\includegraphics[scale=0.5]{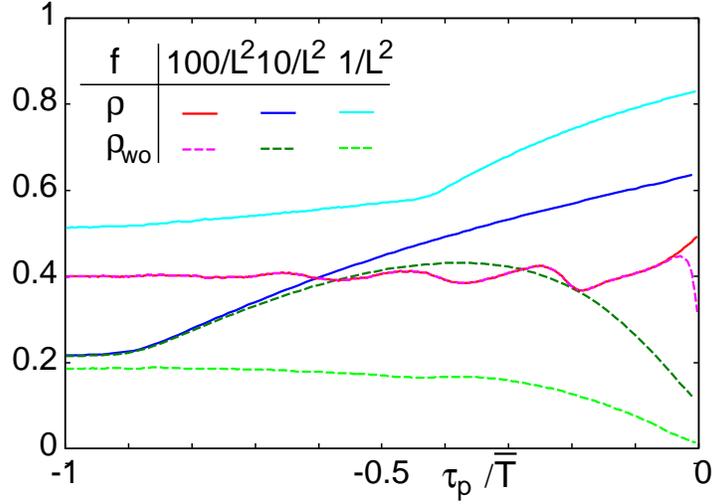}
\caption{Temporal change of the two kinds of stressed ratio: $\rho$ and $\rho_{\rm wo}$.}
 \end{center}
\end{figure}
\begin{figure}
\begin{center}
\includegraphics[scale=0.4]{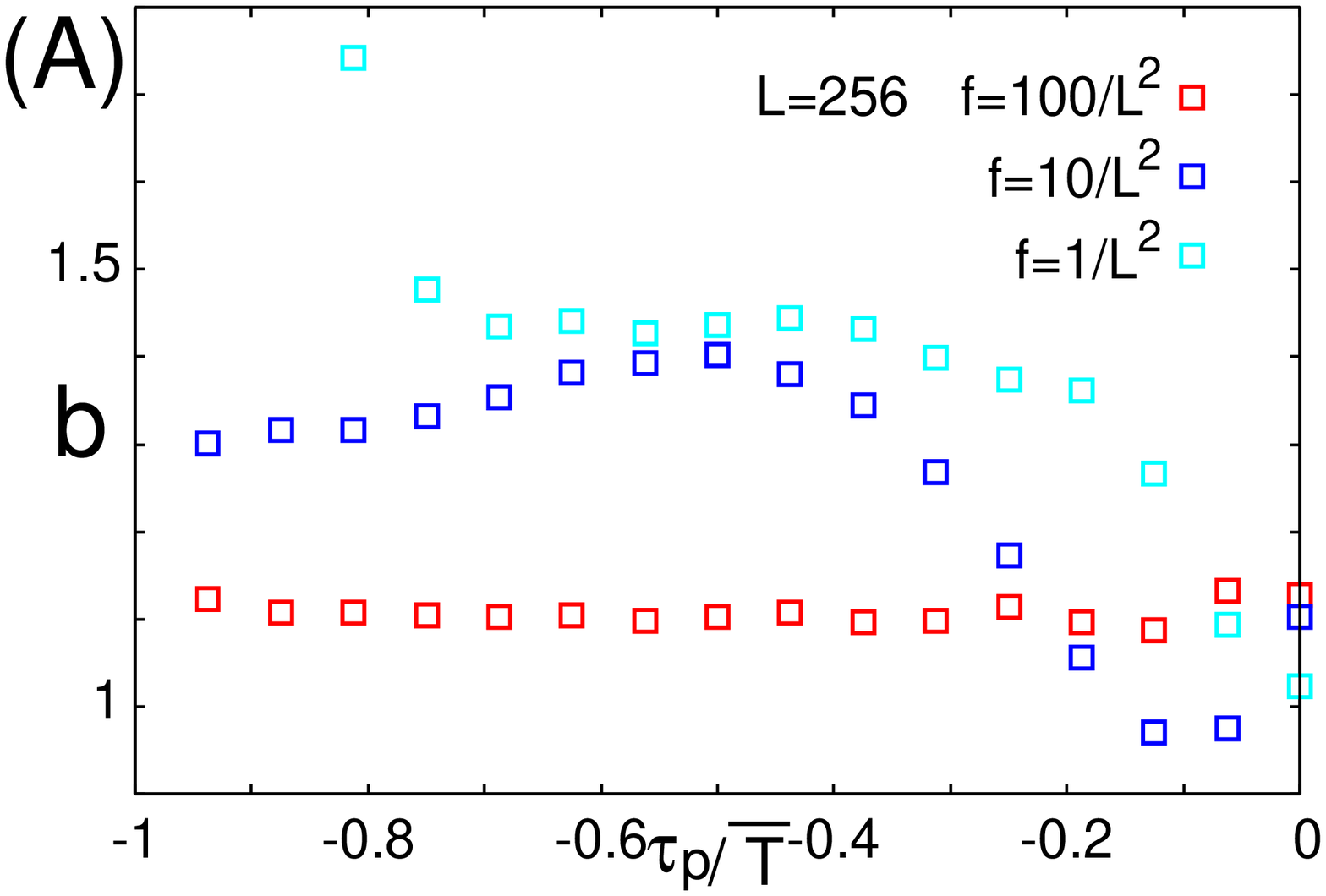}
\includegraphics[scale=0.4]{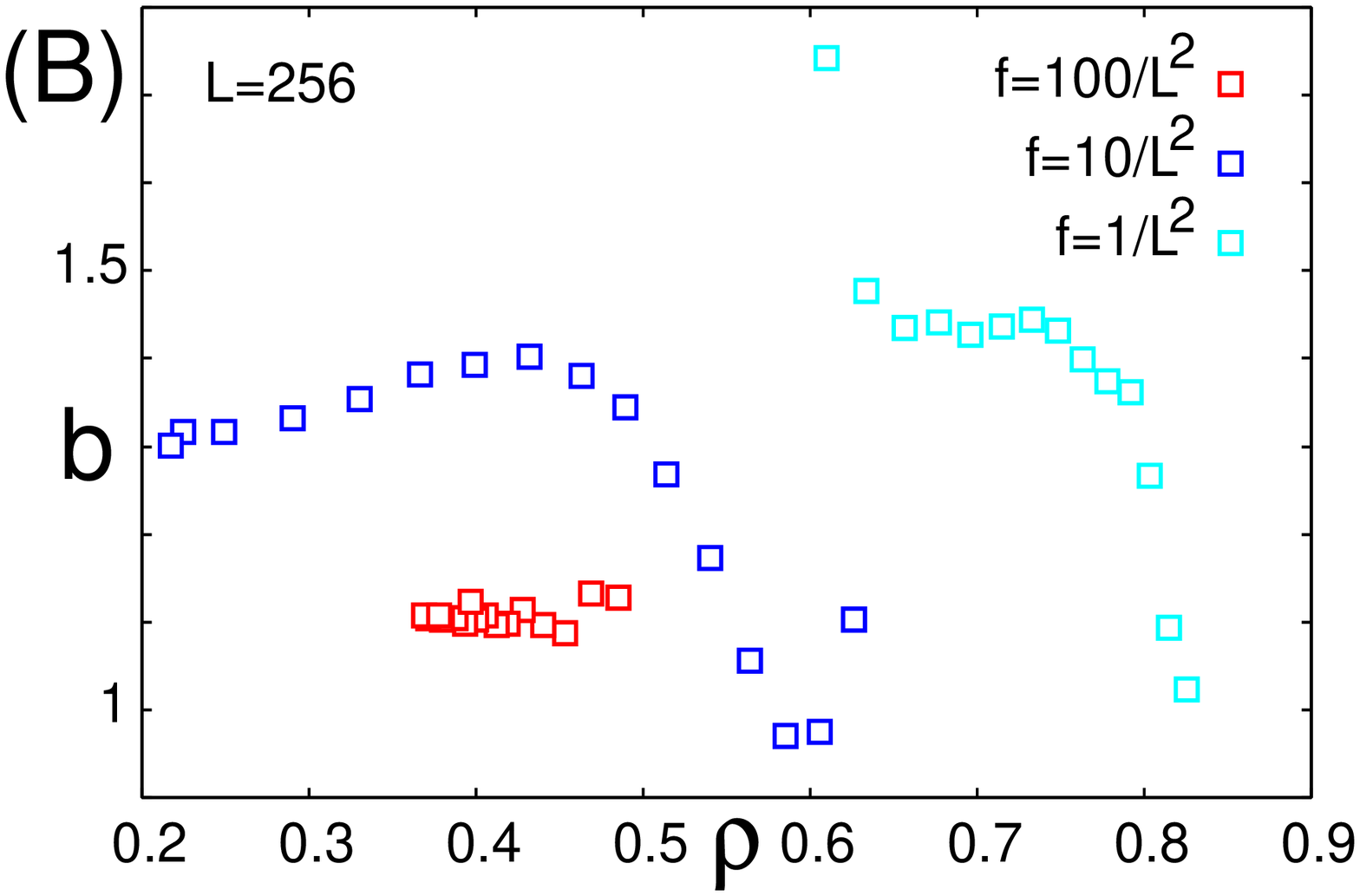}
\caption{(A) Temporal change of b-value with scaled time $\tau_p/\bar{T}$ for
  $f=100/L^2$(red), $f=10/L^2$(blue) and $f=1/L^2$(light blue).
(B) Relation between the $b$-value and the stressed fraction $\rho$.
In these two panels, data are the same as those in FIG.~4. }
 \end{center}
\end{figure}
\end{document}